\documentclass[aps,twocolumn,amssymb,prc]{revtex4-1}
\date{\today}
\usepackage{graphicx}
\usepackage{color}

\usepackage{array}
\usepackage{graphicx}
\usepackage{amssymb}
\usepackage{longtable}

\usepackage{bm}
\begin{document}
\bibliographystyle{h-physrev}

\title{Theoretical uncertainties within the Duflo-Zuker mass model}
\author{Chong Qi}
\thanks{Email: chongq@kth.se}
\affiliation{Department of Physics, Royal Institute of Technology (KTH), Alba Nova University Center,
SE-10691 Stockholm, Sweden}\date{\today}

\begin{abstract}
It is becoming increasingly important to understand the uncertainties of nuclear mass model calculations and their limitations when extrapolating to driplines.
In this paper we evaluate the parameter uncertainties the Duflo-Zuker (DZ) shell model mass formulae by fitting to the latest experimental mass compilation AME2012 and analyze the propagation of the uncertainties in binding energy calculations when extrapolated to driplines.
The parameter uncertainties and uncertain propagations are evaluated with the help of the covariance matrix thus derived. Large deviations from the extrapolations of AME2012 are seen in superheavy nuclei. A simplified version of the DZ model (DZ19) with much smaller uncertainties than that of DZ33 is proposed. Calculations are compared with results from other mass formulae. Systematics on the uncertainty propagation as well as the positions of the driplines are also presented.
The DZ mass formulae are shown to be well defined with good extrapolation properties and rather small uncertainties, even though some of the parameters of the full DZ33 model cannot be fully determined by fitting to available experimental data.
\end{abstract}

\maketitle

\section{Introduction}
The nuclear mass (or the binding energy)  has a rich and respectable history of study, both experimentally and theoretically, and is still a very active field of research \cite{Lunney,Audi2003,ISI:000314891000003,Kan12}. It played an important role in our understanding of the nuclear pairing \cite{Ber09,Satu98,PhysRevC.88.064329}, proton-neutron correlation \cite{Macc00,Satu01,Chas07,Qi2012436,Qi11} as well as the Wigner effect. Great progress has been made in 
measuring the mass of exotic nuclei, however, theoretical models are
still necessary to predict the masses of nuclei far from
stability. In particular, nearly all the nuclear masses along the
predicted astrophysical r-process path are experimentally unknown and
rely on theoretically estimations \cite{Grawe2007,Lan2013,Ber2010}. 
The well-known liquid drop model  is still widely applied, with various changes, in nuclear mass calculations (see, e.g., Refs. \cite{Kir2008,Wang2011,Wang2014,Men2008,Men2008a,Bar2012,Bha2010,Bha2012,Mor2012,Moller1995,Mol12,Sob2014}), which can reproduce all known experimental data
to an average precision of around 0.5 MeV or even less. Large-scale calculations have been done with the Skyrme  \cite{Bender2003,Kor2010,Kor2012,Erl2012,Gor10,Was2012} and Gogny \cite{Gor2009,Del2010}  forces and relativistic mean field approaches \cite{Agr2012}.
When including phenomenological correction terms, structure model calculations within above framework can reproduce the binding energies
in a way comparable with those of empirical mass formulas.

One of the most important and challenging frontiers of
nuclear physics is the study of nuclei at the limit
of stability, especially neutron-rich nuclei with weakly
bound neutrons. However, even though the models mentioned above agree very well within the region of known
masses, they can differ by upto tens of MeV in those unstable regions where the masses are unknown.  In this context, in recent years there has been increased interest in understanding the uncertainties and limitations of the
nuclear mass models \cite{Rei2010,Dob2014,Agb2014,Kor2013,Gao2013,Fat2011,Gor2014,Ber2013}.
The theoretical uncertainties of the relativistic Hartree-Bogoliubov approach are studied in Ref. \cite{Agb2014} in which it is suggested that the uncertainty is mainly due to the poorly controlled isovector channel of the interaction and the inaccurate description of the single-particle energies. Further optimization of the Skyrme Hartree-Fock-Bogoliubov theory is done recently by fitting to both global nuclear properties and the single-particle level splittings \cite{Kor2013}. It is noticed that the constraint on the Skyrme parameters is only marginally improved. As can be seen in Tables III and IV of Ref. \cite{Kor2013}, some of the Skyrme parameters still show pretty large uncertainty which cannot be pinned done by present optimization.
Moreover, as shown in Ref. \cite{Gao2013}, the standard error of the Skyrme functional shows a divergent behavior when goes towards neutron-rich nuclei.

The purpose of this work is to explore the uncertainties of the microscopically inspired Duflo and Zuker (DZ)  shell model mass formula \cite{dz,Duflo1995} by confronting it to the latest AME2012 mass compilation \cite{ISI:000314891000003}, hoping that a model with well controlled uncertainties will lead to more reliable predictions  when extrapolating to the unknown regions. 
We choose the DZ mass model particularly because it is constructed starting from a shell-model monopole Hamiltonian and is based on the sequential filling of a pre-assumed shell structure. 
It is expected that the extended high-accurate mass data may help
with the further improvement of the model and to a better understanding of its driving terms as well as its limitations. 
Moreover, a topic of particular interest is the evolution of the shell structure. That is, the
magic number may change dramatically depending on
the $N/Z$ ratio when we move towards the particle drip
lines~ (See, e.g., Refs. \cite{Sor08,Qi13}).
We hope that further insight on the shell structure and stability of those nuclei may be obtained by comparing their masses with the prediction of the DZ model.

A short introduction to the DZ model is given in Section II. The optimization of the parameters of full DZ model as well as its simple version is given in Section III.
A detailed comparison between the DZ model and several other mass models and extrapolations is presented. A simplified DZ model with only 19 terms is also proposed. The uncertainty propagations of the DZ models are evaluated in Section IV. 
The positions of the proton and neutron driplines predicted by the DZ models are given in Section V. 

\section{The DZ mass model}

The DZ mass model is constructed starting from a shell-model monopole Hamiltonian as
\begin{equation}
BE=<H_m> - E_C-E_{sym} + E_P,
\end{equation}
where the monopole Hamiltonian represents an averaged
mean field extracted from the interacting shell model. $E_C$ and $E_P$ are the Coulomb and pairing energies. The symmetry energy is defined by a two-term expression as
\begin{equation}
E_{sym}=a_{sym}\frac{T(T+1)}{A\rho}-a_{ssym}\frac{T(T+1)}{A^{4/3}\rho^2},
\end{equation}
where $\rho=A^{1/3}[1-\left(\frac{T}{A}\right)^2]^2$ is a scaling factor and the first and second terms are the symmetry energy and surface symmetry energy, respectively.
The monopole Hamiltonian is defined as
\begin{equation}\label{dz}
H_m= H_M+H_s + H_d,
\end{equation}
where $H_M$ is a macroscopic term involving all nucleons. 
The microscopic spherical $H_s$ and deformed $H_d$ parts of the Hamiltonian take into account the residual correlation of 
valence nucleons in the open shell. 
The master Hamiltonian $M+T$ and its surface term $(M+T)/\rho$ resemble the volume and surface energies of the liquid drop model but contain a strong HO shell effect. This is compensated by other terms.

In the spherical case of the DZ mass model, the binding energy is calculated by assuming normal filling of the proton and neutron orbitals. Deformation is simply defined as the promotion of four protons and four neutrons to the next major shell. For deformed nuclei, the quadrupole correlation energy thus gained through $H_d$ may eventually offset the loss of spherical monopole energy. In practice,  for nuclei with $Z>50$, both sphere and deformation calculations are done and the lower binding energy is kept.

\subsection{The simplified ten-term DZ mass model  (DZ10)} 
There are several different versions of the DZ model available \cite{Duflo1995}.
The original DZ10 model contains four macroscopic terms, including the Coulomb energy, symmetry and surface symmetry energies and the pairing energy, and six monopole terms. A detailed study on the role played by different terms of the DZ10 model was presented in Ref. \cite{Mendoza-Temis2010,Bar2012a}.
There are also two correction terms to the surface symmetry energy and the pairing energy of the forms $T(T-1/2)/A\rho^4$ (denoted as 'Wigner energy' in Ref. \cite{dz}) and $2T/A\rho$. These two terms, which have negligible influence on the global description of the binding energy, are not included in our following studies (see, also, Ref. \cite{Qi2012436}).

The DZ10 mass model contains a rather sophisticated $S$ term which is added to the master term $M$. The competition between $M$ and $S$ is responsible for changing the shell structure from HO to spin-orbital ones with $N (Z) = 28, 50, 82, 126$ and 184 \cite{Zuker2008}.

The spherical term $H_s$ in Eq.~(\ref{dz}) is given as
\begin{equation}\label{dz_sph}
<H_{s}>= \frac{1}{\rho} \left[a_{s} {S_{3}} +
b_s \frac{S_{3}}{\rho} +c_sS_{4}\right],
\end{equation}
where  $a_s$, $b_s$ and $c_s$ are constants to be determined. 
The expectation value of the Hamiltonian $H_s$ is calculated by assuming the normal filling scheme of nucleons.
The deformed Hamiltonian $H_d$ can be constructed in a way  similar to $S_4$ but takes into account the effect of the promotion of four valence nucleons to
the next shell~\cite{Duflo1995}.

The pairing energy is given as
\begin{equation}\label{pair}
E_P=a_p\frac{2-v}{\rho},
\end{equation}
where $v$ denotes the seniority of the nucleus. In the present work the seniority quantum number is assumed to be zero for the ground states of even-even nuclei, one for those of odd-$A$ nuclei and two for odd-odd nuclei with isospin $T=|N-Z|/2$ \cite{Qi2012436}. The seniority is assumed to be zero for the $T=1$ ground states of odd-odd $N=Z$ nuclei.

\subsection{The full DZ mass model} 
The full DZ mass model contains 28 monopole terms, which include
FM+(which equals to M+T), fm+,
FS+,  fs+, FS-, fs-,
FC+, fc+, PM+,  pm+, 
PS+, ps+, PS-, ps-, 
S3,  s3,  SQ-, sq-,
D3,  d3,  QQ+,  qq+,  
D0,  d0, QQ-, qq-,  
SS,  ss,
as well as the Coulomb energy, symmetry energy, surface symmetry energy and two pairing terms. There are 33 terms in total (DZ33).
Detailed explanation of the different monopole terms may be found in Ref. \cite{Duflo1995} and will not be repeated here for simplicity. A short explanation of the DZ model can also be found in Ref. \cite{bertsch}.
We just point out that the monopole terms can be separated into two groups with the same numbers of terms: The volume terms (capital letters) and the surface terms (small letters). Among the 28 monopole terms, S3,  s3,  SQ- and sq- are spherical terms while D3,  d3,  QQ+,  qq+,  
D0,  d0, QQ- and qq- are deformed terms. It should be mentioned that, in the original paper \cite{Duflo1995}, only 28 terms (among which there are 24 monopole terms ) are considered. Their parameters are fitted to the AME1993 mass table and compared to that determined by fitting to the older AME1983 compilation. A 31-term version (by excluding d3 and qq-) can be found in Ref. \cite{dz}.
Recent calculations with the DZ33 model may also be found in Refs. \cite{Men2008,Kir2012}.

It should be mentioned that DZ10 is not a simplified version of the DZ33 model. They contain different monopole terms. In particular, the $S$ term is only present in the DZ10 model.
The shell structure is assumed to be the same in both models.

\section{Optimizations and calculations}
The parameters of the DZ10 and DZ33 models are optimized by fitting the latest experimental data as given in Ref. \cite{ISI:000314891000003}. The least square fitting procedure used in this work is the same as in Refs. \cite{Qi12,Qi13}. All data included in the fitting are considered with the same weight for simplicity. As we will show below, some of the parameters of the DZ33 are strongly correlated. A simplified DZ model with only 19 terms will be proposed guided by such a correlation.

\subsection{The DZ10 mass model}

\begin{table}
  \centering
  \caption{The coefficients (in MeV) of the DZ10 mass model determined by fitting to different sets of binding energies: (I) experimental measured binding energies with errors smaller than 100 keV; (II) all measured binding energies, and (III) all experimental and extrapolated binding energies from Refs. \cite{Audi2003,ISI:000314891000003}. The corresponding standard root mean square deviations from experimental data are given at the end (in MeV).}\label{table}
\begin{ruledtabular}
  \begin{tabular}{cccc}
Term& I&II&III\\
\hline 
$E_C$ & 0.705  $\pm$     0.00068 &       0.705  $\pm$     0.00064 &       0.707  $\pm$     0.00045\\
$a_{sym}$ & 148.339  $\pm$     0.290 &     148.429  $\pm$     0.267 &     149.525  $\pm$     0.180\\
$a_{ssym}$ & 203.266  $\pm$     1.335 &     203.749  $\pm$     1.161 &     207.349  $\pm$     0.744\\
$a_P$ &   5.396  $\pm$     0.137 &       5.406  $\pm$     0.131 &       5.236  $\pm$     0.111\\
\hline
$M+T+S$& 17.742  $\pm$     0.013  &      17.738  $\pm$     0.012 &      17.759  $\pm$     0.009\\
$(M+T)/\rho$  &  16.213  $\pm$     0.044 &      16.203  $\pm$     0.042 &      16.236  $\pm$     0.031\\
$S_3$ &   0.460  $\pm$     0.011 &       0.465  $\pm$     0.010 &       0.372  $\pm$     0.007\\
$S_3/\rho$ &   2.079  $\pm$     0.057 &       2.113  $\pm$     0.054 &       1.671  $\pm$     0.035\\
 $S_4$&   0.021  $\pm$     0.00029 &       0.021  $\pm$     0.00029 &       0.019  $\pm$     0.00023\\
$H_d$ &  41.282  $\pm$     0.391 &      41.448  $\pm$     0.382 &      42.941  $\pm$     0.281\\
\hline
$\sigma$(I) & 0.537& 0.538&0.651\\
$\sigma$(II)  &0.573&  0.572&0.677\\
$\sigma$(III) &1.061& 1.061&0.761\\
   \end{tabular}
  \end{ruledtabular}
\end{table}

\begin{figure}[htdp]
\centerline{\includegraphics[width=0.48\textwidth]{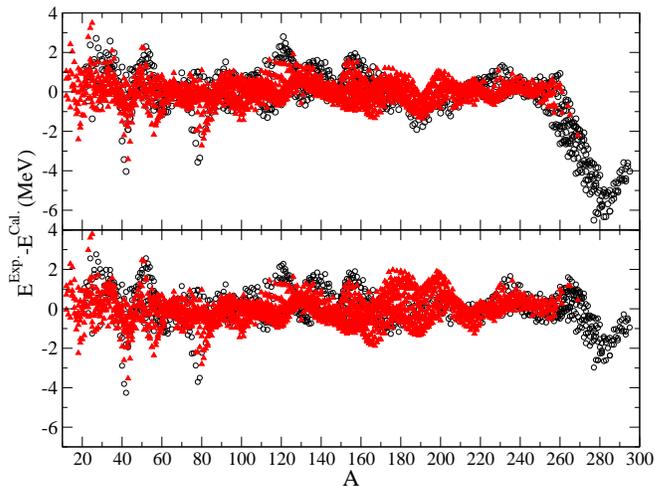} }
\caption{(Color online) Upper: Difference between DZ10 mass model calculations with the parameter set II from Table \ref{table} and experimental data (solid red) as well as extrapolated values (black open). Lower: Same as the upper panel but corresponds to binding energy calculations with the parameter set III. \label{dz10diff}}
\end{figure}

Firstly we determine the parameters of the DZ10 mass model by fitting to known experimental binding energies of nuclei  with $Z\geq6$ and $A\geq12$. Two calculations are done by fitting to experimental data that have errors smaller than 100keV (I) and to all experimental data  (II) listed in Ref. \cite{ISI:000314891000003}. We include in total 2195 and
2325 masses in above two fittings, respectively. The fitted coefficients and the corresponding uncertainties are shown in Table \ref{table}. As can be seen from the Table, these two fittings give practically the same results, which indicates that the DZ10 model has a good extrapolation property. The difference between  experimental data and calculations with the parameter set II is plotted in the upper panel of Fig. \ref{dz10diff}.  It is seen that most experimental data can be well reproduced within an error of around 1 MeV. Only in a few cases the errors are larger than 2.5 MeV. These correspond to neutron rich nuclei $^{23-25}$O, $^{43}$P, $^{44}$S and $^{80}$Zn. It may be related to the fact that the shell structure in these nuclei are rather different than the traditional SO ones employed in the DZ model, which cannot be fully compensated by the spherical monopole Hamiltonian. For heavy and superheavy nuclei, the biggest difference is seen in the case $^{269}$Ds.

A good agreement between the older AME1993 extrapolation and the 28-term DZ model was obtained in Ref. \cite{Duflo1995}.
In the upper panel of Fig. \ref{dz10diff} we also give the deviation of the DZ10 mass calculations with the extrapolation values derived in AME2012 by Wang, Audi and collaborators \cite{ISI:000314891000003}. For light and medium-mass nuclei, the biggest differences ($>2.5$ MeV) appear in $^{27}$O, $^{34}$Na, $^{41}$Al, $^{42}$Si, $^{77,78}$Ni, $^{79}$Cu and $^{121}$Pr. The root-mean-square error between the DZ10 mass calculation and all binding energies from Ref. \cite{ISI:000314891000003} is significantly enlarged to around 1 MeV. This is mainly due to the large deviation seen in the superheavy region for nuclei with $Z\sim$ 110 and $N\sim$ 170. 

It may be interesting to see whether such a large and systematic difference between the AME2012 extrapolation and the DZ10 mass model in the superheavy region is intrinsic due to the different natures of the models used or can be absorbed by renormalizing the parameters. To illustrate this point, in the fourth column of Table \ref{table} we refitted the parameters of the DZ10 model to both experimental and extrapolated binding energies given in Ref. \cite{ISI:000314891000003}. We have considered in total 3229 masses with the same weight in the fitting among which nearly 1/3 are extrapolated values. It can be seen from the table that the coefficients of the spherical  Hamiltonian $H_s$ are noticeably modified in comparison with those of the parameter sets I and II. As a result, the mean deviation from the experimental data is increased by around 100 keV. However, the systematic deviation in the superheavy region persists, as can be seen from Fig. \ref{dz10diff}. 
In this context, one may expect that further mass measurements in this region will provide a critical test to the DZ mass model.

\begin{figure}[htdp]
\centerline{\includegraphics[width=0.48\textwidth]{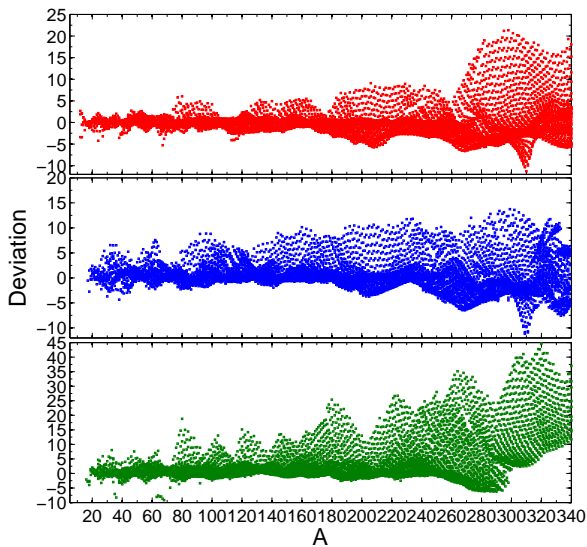} }
\vspace{-2cm}
\caption{(Color online) Differences (in MeV) between DZ10 mass model calculations with the parameter set II from Table \ref{table} and the macroscopic-microscopic liquid drop model calculation of Ref. \cite{Wang2011} (upper), the finite-range droplet model \cite{Moller1995} (middle) and the HFB-21 model \cite{Gor10} (bottom).\label{dz10wang}}
\end{figure}

In Fig. \ref{dz10wang} we also plotted the deviations of the DZ10 calculation from the systematic calculations with the macroscopic-microscopic liquid drop model  \cite{Wang2011}, finite-range droplet model  \cite{Moller1995}  and self-consistent mean-field model calculations with the Skyrme force \cite{Gor10}. With the inclusion of several correction terms, Ref. \cite{Wang2011} can reproduce experimental data within an average deviation of 336 keV. The latest Bsk mass model \cite{Gor10} contain 30 parameters and can reproduce experimental data by around 550 keV. However, large deviations between the DZ10 model and above three models are seen in dripline nuclei and in superheavy nuclei, which indicate that these models have very different extrapolation properties.

\subsection{The full DZ mass model (DZ33)}

\begin{figure}[htdp]
\centerline{\includegraphics[width=0.48\textwidth]{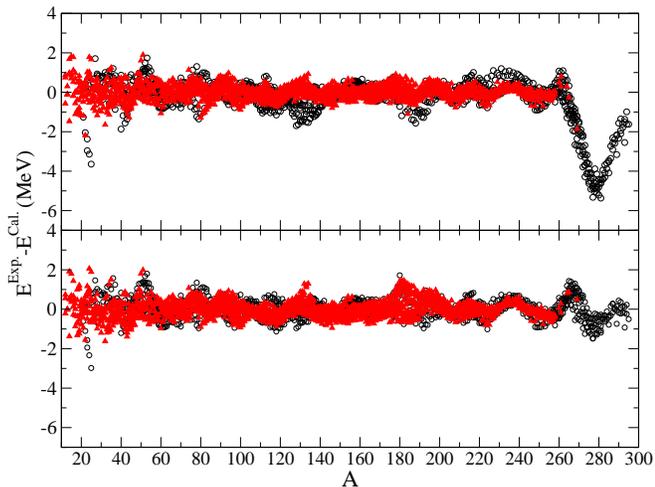} }
\caption{(Color online) Upper: Difference between DZ33 mass model calculations with the parameter set II from Table \ref{table2} and experimental data (solid red) as well as extrapolated values (black open). Lower: Same as the upper panel but corresponds to binding energy calculations with the parameter set III.\label{dz33diff}}
\end{figure}

As in the DZ10 case, we fit the coefficients of the full DZ mass model to experimental binding energies with errors $\leq100$ keV (I), all experimental data (II) and all binding energies listed in Ref. \cite{ISI:000314891000003} using the least-square criterion. The results are given in Table \ref{table2}. Calculations with the parameter sets I and II give quite similar results. Both can reproduce experimental data within an averaged error of around 370 keV. The deviation of calculations with parameter set II from experimental data is plotted in Fig. \ref{dz33diff}. Only in a few cases the difference is larger than 1.5 MeV. These correspond to nuclei $^{22}$N, $^{16, 24,25}$O, $^{14}$F, $^{32}$S, $^{50}$K, $^{51}$Ca around shell closures and the superheavy nucleus $^{269}$Ds. 

\begin{table}
  \centering
  \caption{Same as Table \ref{table} but for the fitted coefficients of the full DZ33 mass model.}\label{table2}
\begin{ruledtabular}
  \begin{tabular}{cccc}
Term& I&II&III\\
\hline
$E_C$ &       0.702  $\pm$     0.002 &       0.702 $\pm$     0.002 &       0.701 $\pm$     0.001\\
$a_{sym}$  &     149.744  $\pm$     0.721 &     149.993 $\pm$     0.625 &     149.286 $\pm$     0.342\\
$a_{ssym}$ &     209.351  $\pm$     2.731 &     210.699 $\pm$     2.274 &     207.180 $\pm$     1.290\\
$a_P$(I) &       6.195  $\pm$     0.162 &       6.259 $\pm$     0.156 &       6.196 $\pm$     0.135\\
$a_P$(II) &       9.741  $\pm$     3.594 &       8.995 $\pm$     3.399 &       8.079 $\pm$     3.045\\
\hline
FM+ &      18.382  $\pm$     0.183 &      18.322 $\pm$     0.173 &      18.623 $\pm$     0.119\\
fm+ &      14.965  $\pm$     0.911 &      14.522 $\pm$     0.861 &      15.813 $\pm$     0.618\\
FS+  &       5.160  $\pm$     0.456 &       5.385 $\pm$     0.426 &       3.417 $\pm$     0.290\\
fs+  &      23.679  $\pm$     2.429 &      24.823 $\pm$     2.265 &      14.580 $\pm$     1.625\\
FS-  &       1.692  $\pm$     0.152 &       1.747 $\pm$     0.145 &       1.357 $\pm$     0.104\\
fs- &       7.708  $\pm$     0.726 &       7.913 $\pm$     0.693 &       6.417 $\pm$     0.525\\
FC+  &      -4.900  $\pm$     1.518 &      -5.486 $\pm$     1.430 &      -1.948 $\pm$     1.064\\
fc+  &     -41.955  $\pm$     4.159 &     -42.897 $\pm$     3.956 &     -28.526 $\pm$     2.968\\
PM+  &      -0.453  $\pm$     0.081 &      -0.436 $\pm$     0.077 &      -0.644 $\pm$     0.053\\
pm+  &      -0.116  $\pm$     0.400 &       0.039 $\pm$     0.379 &      -0.819 $\pm$     0.273\\
PS+  &      -0.842  $\pm$     0.084 &      -0.872 $\pm$     0.079 &      -0.535 $\pm$     0.055\\
ps+ &      -4.264  $\pm$     0.469 &      -4.425 $\pm$     0.439 &      -2.553 $\pm$     0.312\\
PS-  &      -0.106  $\pm$     0.025 &      -0.127 $\pm$     0.022 &      -0.136 $\pm$     0.017\\
ps-  &      -0.562  $\pm$     0.121 &      -0.668 $\pm$     0.107 &      -0.694 $\pm$     0.081\\
S3   &       0.427  $\pm$     0.024 &       0.436 $\pm$     0.023 &       0.455 $\pm$     0.017\\
s3   &       1.964  $\pm$     0.115 &       2.007 $\pm$     0.111 &       2.048 $\pm$     0.082\\
SQ-  &       0.334  $\pm$     0.040 &       0.346 $\pm$     0.038 &       0.339 $\pm$     0.025\\
sq- &       1.320  $\pm$     0.221 &       1.401 $\pm$     0.212 &       1.340 $\pm$     0.146\\
D3   &      -0.081  $\pm$     2.514 &       1.286 $\pm$     2.515 &       8.128 $\pm$     1.864\\
d3   &       8.205  $\pm$    14.560 &      16.288 $\pm$    14.571 &      50.351 $\pm$    10.915\\
QQ+  &       6.150  $\pm$     6.190 &       5.795 $\pm$     6.263 &      -1.829 $\pm$     5.342\\
qq+  &       5.322  $\pm$    33.201 &       1.739 $\pm$    33.829 &     -39.814 $\pm$    30.640\\
D0   &     -33.197  $\pm$     2.131 &     -33.496 $\pm$     2.105 &     -23.467 $\pm$     1.330\\
d0 &    -158.619  $\pm$    10.380 &    -159.670 $\pm$    10.249 &    -108.181 $\pm$     7.096\\
QQ-  &      -2.777  $\pm$     6.405 &      -2.035 $\pm$     6.486 &      12.818 $\pm$     5.162\\
qq-  &     -24.519  $\pm$    34.650 &     -18.726 $\pm$    35.331 &      65.004 $\pm$    29.631\\
SS    &       1.152  $\pm$     0.215 &       1.227 $\pm$     0.204 &       1.501 $\pm$     0.148\\
ss &       3.728  $\pm$     0.919 &       4.064 $\pm$     0.862 &       4.310 $\pm$     0.624\\

\hline
$\sigma$(I)& 0.356& 0.358&0.418\\
$\sigma$(II)  &0.377&  0.374&0.430\\
$\sigma$(III)&0.769& 0.770&0.463\\
   \end{tabular}
  \end{ruledtabular}
\end{table}

As in the case of DZ10, both calculations with the parameter sets I and II show large deviations from the extrapolations of Ref. \cite{ISI:000314891000003} in the superheavy region with $A\sim$ 270. However, unlike that in DZ10, it looks as if the deviation partially disappear if we refit the parameters of the DZ33 model to the extrapolations.
Calculations with the parameter set III can still reproduce quite well the experimental data with an average error 430 keV.

\begin{figure}[htdp]
\centerline{\includegraphics[width=0.48\textwidth]{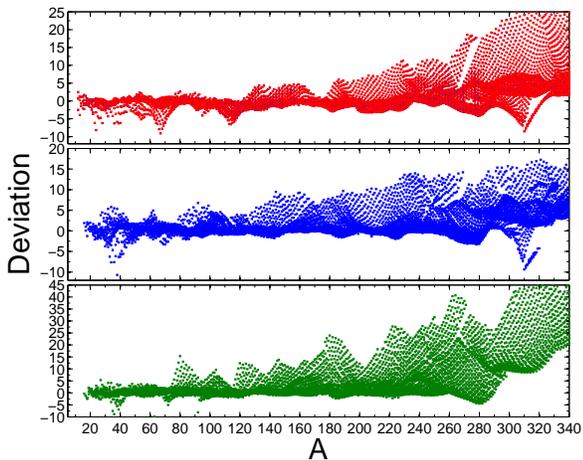} }
\caption{(Color online) Same as Fig. \ref{dz10wang} but for calculations with the DZ33 model.\label{dz33diffwang}}
\end{figure}

The differences between the DZ33 calculations and those with the macroscopic-microscopic liquid drop model  \cite{Wang2011}, finite-range droplet model  \cite{Moller1995}  and self-consistent mean-field model calculations with the Skyrme force \cite{Gor10} are plotted in Fig. \ref{dz33diffwang}. As can be seen from the figure, the differences show a pattern that is pretty similar to that of DZ10. This is related to the fact that DZ33 and DZ10 give pretty similar results for all nuclei upto the driplines.

\subsection{Parameter uncertainties and the DZ model with reduced terms (DZ19)}

As can be seen from Table II, in quite a few cases the parameters of the DZ33 mass model show pretty large uncertainties, which indicates that those monopole terms cannot be well determined by fitting to available experimental data. These parameter uncertainties may affect the prediction power of the DZ33 model for nuclei in the unknown regions. We are motivated to find a simplified version of the DZ mass model based on the correlation matrix of the parameters. It is hoped that such a simpler mass model would have a comparable performance to that of the full model. We are thus left with a simplified DZ model with only 19 terms (DZ19).
We have only 15 monopole terms left among which there are only six surface terms. We have also a mixed deformation term which combines the surface term d3 and the volume term QQ+. The reason for this is that their parameters show a strong correlation.

\begin{table}
  \centering
  \caption{Same as Table \ref{table} but for the simplified DZ model, DZ19.}\label{table3}
\begin{ruledtabular}
  \begin{tabular}{cccc}
Term& I&II&III\\
\hline 
$E_C$  &       0.705  $\pm$     0.001 &       0.704  $\pm$     0.001 &       0.703 $\pm$     0.001\\
$a_{sym}$  &     149.033  $\pm$     0.468 &     148.988  $\pm$     0.427 &     148.837 $\pm$     0.259\\
$a_{ssym}$ &     202.642  $\pm$     1.848 &     203.222  $\pm$     1.618 &     203.294 $\pm$     0.975\\
$a_P$ &       5.167  $\pm$     0.138 &       5.177  $\pm$     0.133 &       5.114 $\pm$     0.113\\
 \hline
FM+ &      18.552  $\pm$     0.037 &      18.528  $\pm$     0.035 &      18.579 $\pm$     0.025\\
fm+  &      15.120  $\pm$     0.125 &      14.985  $\pm$     0.117 &      14.874 $\pm$     0.078\\
FS  &       1.069  $\pm$     0.040 &       1.066  $\pm$     0.040 &       1.051 $\pm$     0.030\\
fs- &       4.414  $\pm$     0.259 &       4.399  $\pm$     0.253 &       4.634 $\pm$     0.196\\
fc+ &     -10.353  $\pm$     0.753 &      -9.602  $\pm$     0.707 &      -9.171 $\pm$     0.549\\
PM+ &      -0.498  $\pm$     0.011 &      -0.506  $\pm$     0.010 &      -0.547 $\pm$     0.008\\
PS+ &      -0.082  $\pm$     0.008 &      -0.080  $\pm$     0.008 &      -0.072 $\pm$     0.006\\
S3 &       0.515  $\pm$     0.015 &       0.529  $\pm$     0.014 &       0.482 $\pm$     0.010\\
s3 &       2.381  $\pm$     0.069 &       2.441  $\pm$     0.065 &       2.194 $\pm$     0.045\\
SQ- &       0.393  $\pm$     0.022 &       0.389  $\pm$     0.022 &       0.358 $\pm$     0.013\\
sq- &       1.581  $\pm$     0.120 &       1.556  $\pm$     0.116 &       1.376 $\pm$     0.079\\
d3+QQ+ &       7.204  $\pm$     0.169 &       7.223  $\pm$     0.163 &       7.080 $\pm$     0.128\\
D0 &     -21.533  $\pm$     0.699 &     -21.780  $\pm$     0.685 &     -22.702 $\pm$     0.531\\
d0 &     -89.904  $\pm$     3.220 &     -91.242  $\pm$     3.129 &     -96.741 $\pm$     2.428\\
SS &       0.431  $\pm$     0.050 &       0.460  $\pm$     0.048 &       0.460 $\pm$     0.040\\
 \hline
$\sigma$(I)  & 0.425& 0.428&0.486\\
$\sigma$(II)  &0.461&  0.457&0.503\\
$\sigma$(III)&0.711& 0.689&0.551\\
   \end{tabular}
  \end{ruledtabular}
\end{table}

\begin{figure}[htdp]
\centerline{\includegraphics[width=0.48\textwidth]{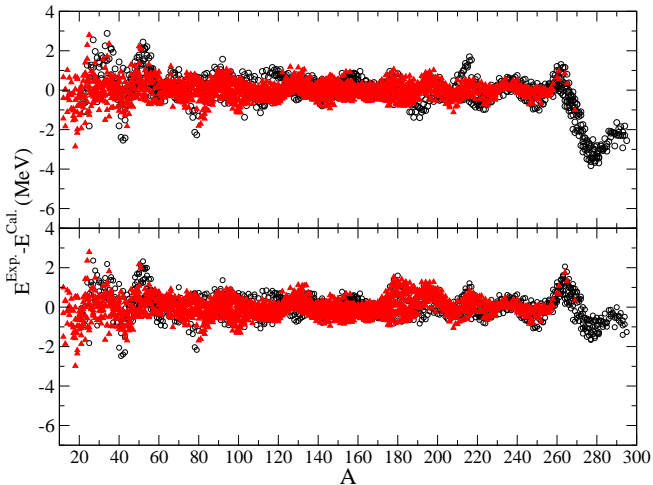} }
\caption{(Color online) Upper: Difference between DZ19 mass model calculations with the parameter set II from Table \ref{table3} and experimental data (solid red) as well as extrapolated values (black open). Lower: Same as the upper panel but corresponds to binding energy calculations with the parameter set III.\label{dz19diff}}
\end{figure}

As in the cases DZ10 and DZ33, we fit the coefficients of the DZ19 mass model to experimental binding energies with errors $\leq100$ keV (I), all experimental data (II) and all binding energies listed in Ref. \cite{ISI:000314891000003} using the least-square criterion. The results are given in Table \ref{table3}. Calculations with the parameter sets I and II give quite similar results. Both can reproduce experimental data within an averaged error of around 430 keV, which is only 70 keV larger than those of the full DZ33 model. The deviation of calculations with parameter set II from experimental data is plotted in Fig. \ref{dz19diff} which shows a pattern that is very close to that of DZ33.

It should be mentioned that for all terms the DZ19 model shows much smaller parameter uncertainties than those of the full DZ33 model.
It may also be interesting to compare the values of the parameters of the DZ19 model with those of the full model given in Tables II and III. It is thus noticed that the coefficients of the Coulomb energy, surface energies as well as the two $M+T$ master terms, which define the bulk properties of the binding energy, remain practically the same. The coefficient of the pairing energy is changed since the second pairing energy term is removed in the DZ19 model. All the four spherical terms are kept even though the values their parameters are somewhat different from the original ones. Moreover, in most cases, the values of the parameters determined by the least square fittings are quite similar to each other.

\section{The uncertainty propagation}
The propagation of parameters uncertainties within the Skyrme HFB approach was considered in Refs. \cite{Gao2013,Gor2014}. Here we evaluate the uncertainty propagation of the DZ mass models mentioned above by applying the same procedure as in Ref. \cite{Gao2013},
\begin{equation}
\sigma^2 (BE)=\sum_{i,j}^N \mathrm{Cov} (x_i,x_j)\frac{\partial BE}{\partial x_i}\frac{\partial BE}{\partial x_j},
\end{equation}
where $BE$ denotes the binding energy, $x_i$ is the parameter and  $N$ denotes the total number of parameters. $\mathrm{Cov} (x_i,x_j)$ is the covariance matrix (see, also, Ref. \cite{Ber2013}) which is derived from the fitting procedure as described in the previous Section.

\begin{figure}[htdp]
\centerline{\includegraphics[width=0.48\textwidth]{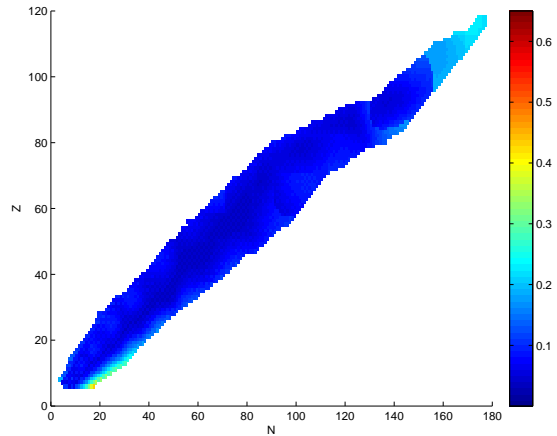} }
\caption{(Color online) The propagation of parameter uncertainties (in MeV) in binding energy calculations within the DZ10 mass model.\label{e10}}
\end{figure}

\begin{figure}[htdp]
\centerline{\includegraphics[width=0.48\textwidth]{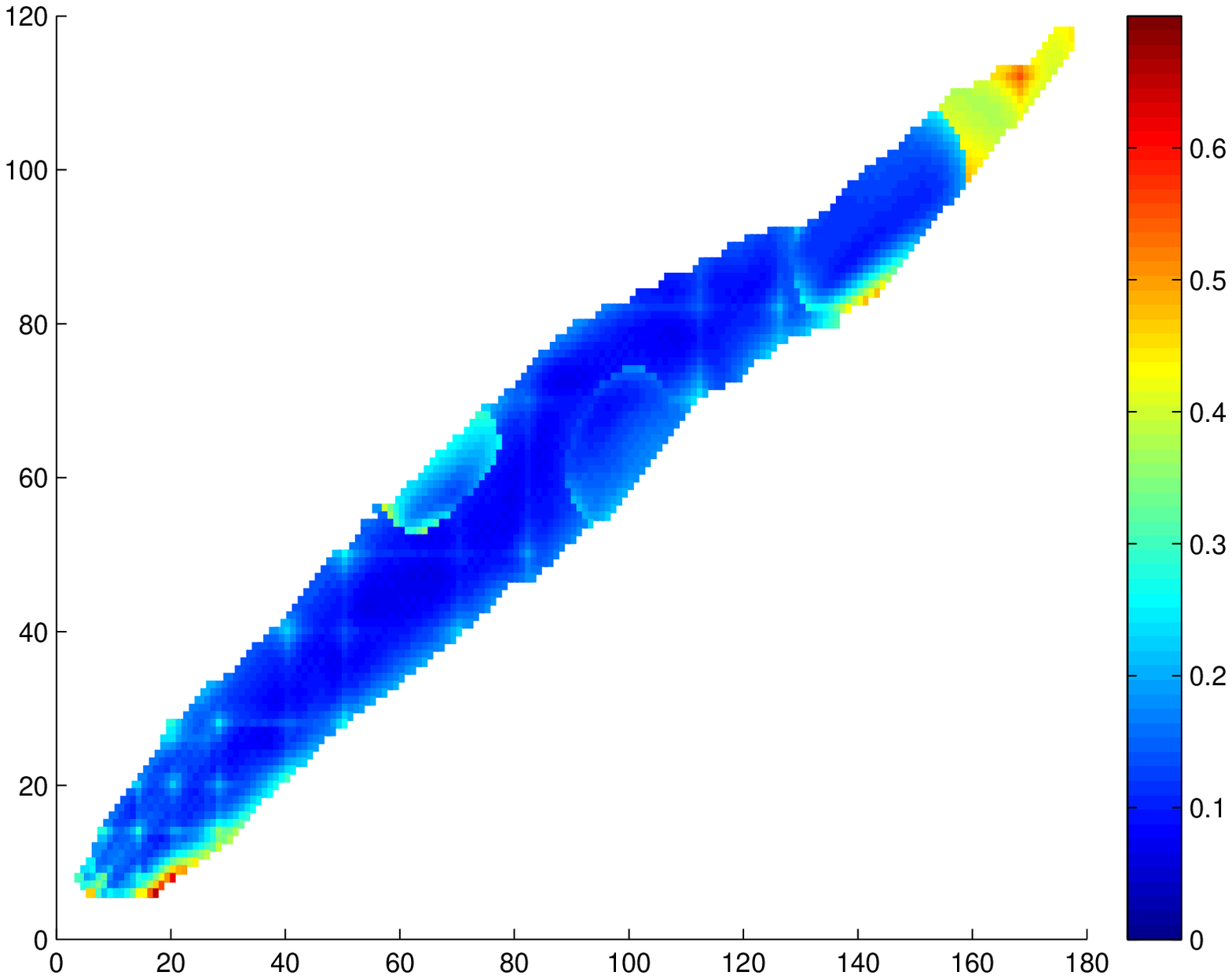} }
\caption{(Color online) Same as Fig. \ref{e10} but for the DZ33 model.\label{e33}}
\end{figure}

\begin{figure}[htdp]
\centerline{\includegraphics[width=0.48\textwidth]{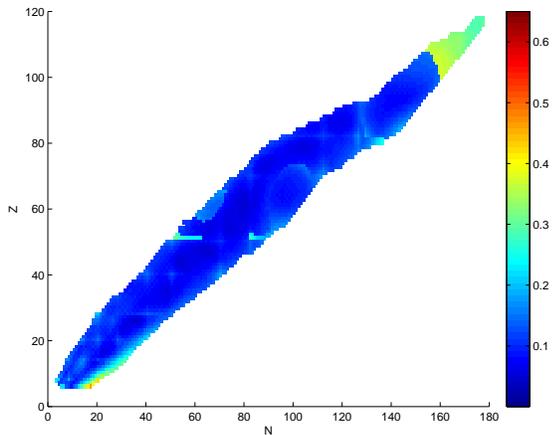} }
\caption{(Color online) Same as Fig. \ref{e10} but for the DZ19 model.\label{e19}}
\end{figure}

The results for the uncertainties in calculated binding energies within the DZ10, DZ33 and DZ19 mass models are given in Figs. \ref{e10}-\ref{e19}. In contrast to that of the Skyrme mean field approach \cite{Gao2013}, all three DZ mass formulae show pretty small uncertainties in binding energy calculations.
For calculations with the DZ10 mass, in most cases the uncertainties are smaller than 200 keV. It is only getting larger in superheavy nuclei and in a few very neutron-rich nuclei. The DZ33 model shows larger uncertainties than those of the DZ10 and DZ19 formulae. This is also related to the fact that some of its parameters are not well defined and show large uncertainties.

\section{The driplines}

\begin{figure}[htdp]
\vspace{1cm}
\centerline{\includegraphics[width=0.48\textwidth]{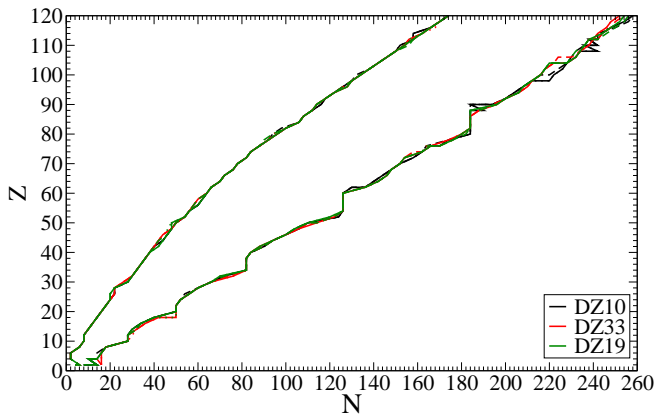} }
\caption{(Color online) The two-neutron and two-proton driplines for even-even nuclei between $2\leq Z\leq 120$ obtained with the DZ10, DZ33 and DZ19 calculations. The parameter sets I (solid), II (dotted) and III (dashed) are used but they cannot be distinguished in the present scale. \label{drip}}
\end{figure}

In Fig. \ref{drip} we plot the calculated two-neutron and two proton driplines for even-even nuclei  between $2\leq Z\leq 120$. The neutron dripline is defined as $S_{2n} (N,Z)>0$ and $S_{2n} (N+2,Z)<0$. The proton dripline is defined in a similar way as $S_{2p} (N,Z)>0$ and $S_{2p} (N-2,Z)<0$. In a few cases one may not have a clear neutron dripline since the nucleus with two more neutrons outside an unbound nucleus may become bound again.
All three DZ mass models are used with the parameters taken from Table I to III. As it can be seen from the figure, the nine calculations give very similar results and thus they can not be clearly distinguished from each other.

In all cases shown in the figure the total number of bound nuclei is calculated to be around 1850. This is lower than the numbers predicted by the non-relativistic mean field approaches given in Ref. \cite{Erl2012}, which are between 1928 (SLy4) and 2333 (SkM*).
Moreover, the positions of the neutron driplines predicted by DZ models with different parameters are pretty similar to each other. This is in contrast to those of the  relativistic \cite{Agb2014,Afa2013} and non-relativistic \cite{Erl2012} mean field approaches, which show rather large spread for calculations with different parameters.

\section{Summary}
Both the ten-term DZ10 model and the full DZ33 model are optimized by fitting to the available experimental data. The parameter uncertainties and uncertain propagations are evaluated with the help of the covariance matrix derived from the fitting. The DZ mass formulae are shown to be well defined with good extrapolation properties and rather small uncertainties. However, some of the parameters of the full DZ33 model cannot be fully determined by fitting to available experimental data and show large uncertainties.
 A simplified version of the DZ model (DZ19) with much smaller uncertainties than that of DZ33 is also proposed. 
 
 Calculations with the DZ mass formulae are compared with results from other mass formulae. Large deviations from the extrapolations of AME2012 are seen in superheavy nuclei around $A=280$ for calculations with both formulae. Systematics on the uncertainty propagation as well as the positions of the driplines are also presented.
The DZ10, DZ33 and DZ19 mass formulae give pretty similar results for the two-neutron and two-proton driplines. The total number of bound even-even nuclei is predicted to be around 1850 in all calculations, which is smaller than those given by recent Skyrme mean field calculations.

By taken into account the fact that most of the parameters of the DZ mass model can already be well defined by fitting to available experimental data, further optimization of the model may be possible by developing an energy functional theorgy from the model based on variational occupations or by having a more sophisticated shell scheme by taking into the evolution of the shell structure in dripline nuclei.

\section*{Acknowledgement}

The author thanks R. Liotta and R. Wyss for stimulating discussions and all their support.
This work was supported by the Swedish Research Council (VR) under grant Nos. 621-2012-3805, and
621-2013-4323.
The
calculations were performed on resources
provided by the Swedish National Infrastructure for Computing (SNIC)
at NSC in Link\"oping.


\begin{thebibliography}{150}
\bibitem{Lunney} D. Lunney, J. M. Pearson and C. Thibault, Rev. Mod. Phys. {\bf 75} 1021  (2003).
\bibitem{Kan12}A. Kankainen, J. \"Ayst\"o, A. Jokinen, J. Phys. G 39, 093101 (2012).
\bibitem{Audi2003}
G.~Audi, A.~Wapstra, and C.~Thibault,
\newblock {Nucl. Phys. A} {\bf {729}}, 337 ({2003}).

\bibitem{ISI:000314891000003}
M. Wang, G. Audi, A. H. Wapstra, F. G. Kondev, M. Maccormick, X. Xu, and B. Pfeiffer, Chin. Phys. C {\bf 36}, 1603 (2012).

\bibitem{Ber09}G. F. Bertsch, C. A. Bertulani, W. Nazarewicz, N. Schunck, and M. V. Stoitsov, 
Phys. Rev. C {\bf79}, 034306 (2009), and references therein.
\bibitem{Satu98}W. Satu{\l}a, J. Dobaczewski, and W. Nazarewicz, Phys. Rev. Lett. {\bf81}, 3599 (1998).
\bibitem[{Hove et~al.(2013)Hove, D. and Jensen, A. S. and Riisager,
  K.}]{PhysRevC.88.064329}
\bibinfo{author}{D.~Hove}, \bibinfo{author}{A.~S. Jensen},
  \bibinfo{author}{K.~Riisager}, \bibinfo{journal}{Phys. Rev. C}
  \bibinfo{volume}{88} (\bibinfo{year}{2013}) \bibinfo{pages}{064329}.

\bibitem{Macc00}A.O. Macchiavelli {\textit et al.}, Phys. Rev. C {\bf61}, 041303(R) (2000).
\bibitem{Satu01}W. Satu{\l}a and R. Wyss, Phys. Rev. Lett. {\bf86}, 4488 (2001); 
Phys. Rev. Lett. {\bf87}, 052504 (2001).
\bibitem{Chas07}R.R. Chasman, Phys. Rev. Lett. {\bf99} (2007) 082501.
\bibitem{Qi2012436}
C.~Qi,
\newblock Phys. Lett. B {\bf 717}, 436  (2012).
\bibitem{Qi11}C. Qi, J. Blomqvist, T. B\"ack, B. Cederwall, A. Johnson, R. J. Liotta, and R. Wyss,
Phys. Rev. C 84, 021301 (2011).

\bibitem{Grawe2007} H. Grawe, K. Langanke, and G. Martinez-Pinedo, Rep. Prog. Phys. {\bf70}, 1525 (2007).
\bibitem{Lan2013} K. Langanke and H. Schatz, Phys. Scr. {\bf T152}, 014011 (2013).
\bibitem{Ber2010} C. A. Bertulani and A. Gade, Phys. Rep. {\bf485}, 195 (2010).

\bibitem{Mol12}P. M\"oller, W.D. Myers, H. Sagawa, and S. Yoshida, 
Phys. Rev. Lett. 108, 052501 (2012).
\bibitem{Kir2008} M. W. Kirson, Nucl. Phys. A {\bf798}, 29 (2008).
\bibitem{Men2008} J. Mendoza-Temis, I. Morales, J. Barea, A. Frank, J. G. Hirsch, J. C. L. Vieyra, P. Van Isacker, and V. Velazquez, Nucl. Phys. A {\bf812}, 28 (2008).
\bibitem{Men2008a} J. Mendoza-Temis, A. Frank, J. G. Hirsch, J. C. L. Vieyra, I. Morales, J. Barea, P. Van Isacker, and V. Velazquez, Nucl. Phys. A {\bf799}, 84 (2008).
\bibitem{Wang2011}M. Liu, N. Wang, Y. G. Deng, and X. Z. Wu, Phys. Rev. C {\bf 84}, 014333 (2011).
\bibitem{Wang2014} N. Wang, M. Liu, X. Wu, and J. Meng, arXiv:1405.2616 (2014).
\bibitem{Bar2012} C. Barbero, J. G. Hirsch, and A. E. Mariano, Nucl. Phys. A {\bf874}, 81 (2012).

\bibitem{Bha2010} A. Bhagwat, X. Vinas, M. Centelles, P. Schuck, and R. Wyss, Phys. Rev. C {\bf81}, 044321 (2010).
\bibitem{Bha2012} A. Bhagwat, X. Vinas, M. Centelles, P. Schuck, and R. Wyss, Phys. Rev. C {\bf86}, 044316 (2012).
\bibitem{Mor2012} L. G. Moretto, P. T. Lake, L. Phair, and J. B. Elliott, Phys. Rev. C {\bf86}, 021303 (2012).
\bibitem{Moller1995} P. M\"oller, J. R. Nix, W. D. Myers, and W. J. Swiatecki, Atomic Data Nucl. Data Tables {\bf59}, 185 (1995).
\bibitem{Sob2014}A. Sobiczewski and Y. A. Litvinov, Phys. Rev. C 89, 024311 (2014).

\bibitem{Bender2003} M. Bender, P. H. Heenen, and P. G. Reinhard, Rev. Mod. Phys. {\bf75}, 121 (2003).
\bibitem{Kor2010} M. Kortelainen, T. Lesinski, J. More, W. Nazarewicz, J. Sarich, N. Schunck, M. V Stoitsov, and S. Wild, Phys. Rev. C {\bf82}, 024313 (2010).
\bibitem{Erl2012} J. Erler, N. Birge, M. Kortelainen, W. Nazarewicz, E. Olsen, A. M. Perhac, and M. Stoitsov, Nature {\bf486}, 509 (2012).
\bibitem{Kor2012} M. Kortelainen, J. McDonnell, W. Nazarewicz, P.-G. Reinhard, J. Sarich, N. Schunck, M. V Stoitsov, and S. M. Wild, Phys. Rev. C {\bf85}, 024304 (2012).
\bibitem{Gor10}S. Goriely, N. Chamel, and J. M. Pearson, Phys. Rev. C 82, 035804 (2010); S. Goriely, N. Chamel, and J. M. Pearson, Phys. Rev. Lett. 102, 152503 (2009);
S. Goriely, N. Chamel, and J. M. Pearson, Phys. Rev. C 88, 024308 (2013); S. Goriely, N. Chamel, and J. M. Pearson, Phys. Rev. C {\bf88}, 061302 (2013).. http://www.astro.ulb.ac.be/pmwiki/Brusslib/Hfb17
\bibitem{Was2012} K. Washiyama, K. Bennaceur, B. Avez, M. Bender, P.-H. Heenen, and V. Hellemans, Phys. Rev. C {\bf86}, 054309 (2012).

\bibitem{Gor2009} S. Goriely, S. Hilaire, M. Girod, and S. Peru, Phys. Rev. Lett. {\bf102}, 242501 (2009).
\bibitem{Del2010} J. P. Delaroche, M. Girod, J. Libert, H. Goutte, S. Hilaire, S. Peru, N. Pillet, and G. F. Bertsch, Phys. Rev. C {\bf81}, 014303 (2010).

\bibitem{Agr2012} B. K. Agrawal, A. Sulaksono, and P.-G. Reinhard, Nucl. Phys. A {\bf882}, 1 (2012).

\bibitem{Dob2014} J. Dobaczewski, W. Nazarewicz, and P.-G. Reinhard, arXiv:1402.4657 (2014).
\bibitem{Rei2010} P. G. Reinhard and W. Nazarewicz, Phys. Rev. C {\bf81}, 051303 (2010).
\bibitem{Agb2014} S. E. Agbemava, A. V. Afanasjev, D. Ray, and P. Ring, arXiv:1404.4901 (2014).
\bibitem{Kor2013} M. Kortelainen, J. McDonnell, W. Nazarewicz, E. Olsen, P.-G. Reinhard, J. Sarich, N. Schunck, S. M. Wild, D. Davesne, J. Erler, and A. Pastore, arXiv:1312.1746 (2013).
\bibitem{Gao2013} Y. Gao, J. Dobaczewski, M. Kortelainen, J. Toivanen, and D. Tarpanov, Phys. Rev. C {\bf87}, 034324 (2013).
\bibitem{Fat2011} F. J. Fattoyev and J. Piekarewicz, Phys. Rev. C {\bf84}, 064302 (2011).
\bibitem{Gor2014} S. Goriely and R. Capote, Phys. Rev. C {\bf89}, 054318 (2014).
\bibitem{Ber2013} M. G. Bertolli, Eur. Phys. J. A {\bf49}, 43 (2013).
\bibitem{dz}
http://amdc.in2p3.fr/web/dz.html.

\bibitem{Duflo1995}
J.~Duflo and A.~Zuker,
\newblock Phys. Rev. C {\bf 52}, R23 (1995).

\bibitem{Sor08}O. Sorlin, M.G. Porquet, Prog. Part. Nucl. Phys. 61, 602 (2008).
\bibitem{Qi13}Z.-X. Xu, and C. Qi, Phys. Lett. B {\bf 724}, 247 (2013).


  \bibitem{Mendoza-Temis2010}J. Mendoza-Temis, J. G. Hirsch, and A. P. Zuker, Nucl. Phys. A {\bf 843}, 14 (2010).
  \bibitem{Bar2012a} C. Barbero, J. G. Hirsch, and A. Mariano, AIP Conf. Proc. {\bf1488}, 162 (2012).

\bibitem{Zuker2008} A. P. Zuker, Rev. Mex. Fis. {\bf54}, 129 (2008).
\bibitem{bertsch}http://www.phys.washington.edu/\verb+~+bertsch/pedlist.html

\bibitem{Kir2012} M. W. Kirson, Nucl. Phys. A {\bf893}, 27 (2012).

\bibitem{Qi12}C. Qi, and Z. X. Xu, Phys. Rev. C {\bf 86}, 044323 (2012).


\bibitem{Afa2013} A. V. Afanasjev, S. E. Agbemava, D. Ray, and P. Ring, Phys. Lett. B {\bf726}, 680 (2013).

\end{thebibliography}

\end{document}